\documentclass[10pt]{article}

\usepackage{amssymb}
\usepackage{color}
\usepackage{epsfig,amssymb,amsfonts,amsmath,graphicx,dsfont,cite,xfrac,multicol}
\usepackage{authblk}
\usepackage{subcaption}
\definecolor{mygray}{gray}{0.5}

\usepackage{cite}
\usepackage[colorlinks=true,linkcolor=blue,citecolor=red]{hyperref}

\newcommand{\be}{\begin{equation}}
\newcommand{\ee}{\end{equation}}
\newcommand{\bea}{\begin{eqnarray}}
\newcommand{\eea}{\end{eqnarray}}

\title{Mathematical-Physics of propagating modes in planar waveguides\\
\textcolor{blue}{\large Accepted in Revista Mexicana de F\'{i}sica E} }

\author[${}$]{J. E. G\'{o}mez-Correa,${}^{a}$ S. E. Balderas-Mata,${}^{b}$ A. Garza-Rivera,${}^{c}$ A. Jaimes-N\'{a}jera,${}^{c,*}$ J. P. Trevino,${}^{d}$ V. Coello,${}^{e}$ J. Rogel-Salazar,${}^{f}$ S. Ch\'{a}vez-Cerda${}^{c}$}

\affil[${}$]{\footnotesize${}^{a}$C\'{a}tedras Conacyt - Centro de Investigaci\'{o}n Cient\'{i}fica y de Educaci\'{o}n Superior de Ensenada, Baja California, Unidad Monterrey, PIIT Apodaca, NL 66629, Mexico.}
\affil[${}$]{
${}^{b}$Universidad de Guadalajara, Departamento de Electr\'onica, C.P. 44840, Guadalajara, Jalisco, Mexico.}
\affil[${}$]{
${}^{c}$Instituto Nacional de Astrof\'isica, \'Optica y Electr\'onica, Coordinaci\'on de \'Optica, Tonantzintla Puebla, 72840, Mexico.}
\affil[${}$]{
${}^{d}$Instituto Tecnol\'ogico y de Estudios Superiores de Monterrey, Departamento de Bioingenier\'ia, Puebla, Puebla, 72453, Mexico.}
\affil[${}$]{
${}^{e}$Centro de Investigaci\'{o}n Cient\'{i}fica y de Educaci\'{o}n Superior de Ensenada, Unidad Monterrey, PIIT Apodaca, NL 66629, Mexico.}
\affil[${}$]{
${}^{f}$Science and Technology Research Institute, School of Physics Astronomy and Mathematics, University of Hertfordshire, Hatfield, Herts., AL10 9AB, UK.}
\affil[${}$]{
${}^{*}$email: ajaimes@inaoep.mx}

\date{}
\begin{document}

\maketitle

\begin{abstract}
In this paper we present a detailed physical analysis of the formation of the propagation transverse modes in planar dielectric waveguides using a mathematical-physics approach. We demonstrate physically that, at the wavelength scale, the pure stationary mode inside planar waveguide is described by the cosine function.  Meanwhile, the sine function yields a quasi-stationary periodic mode.
\end{abstract}

\begin{multicols}{2}

\section{Introduction}
Recently, it has been brought to attention the difference between a mathematical-physics and physical-mathematics approach when dealing with the modeling and finding the mathematical solution of a physical system \cite{JEGO}. The relevance of that difference resides on the fact that the physical properties of the system can be lost or misunderstood if only a pure physical-mathematics approach is used. In general, physics textbooks are based on solving the model of a physical problem by using mathematical methods. Once the solution is obtained, an attempt to provide a physical interpretation may be performed. Unfortunately, in many instances it is a purely mathematical interpretation what is given so that the student obtains the solution in a straightforward way. This form of solving a physical problem is what was referred as the physical-mathematics approach \cite{JEGO,sommerfeld}.

A typical situation of this approach can be found in Optics when studying the modes in planar wave guides and in Quantum Mechanics when analyzing square potential wells \cite{iizuka,Lee,Inan,Ramo,Okamoto,saleh,Chen,Griffiths,Merzbacher,Campi}.

On the other side, within mathematical-physics approach it has been demonstrated the importance of the inclusion of the Neumann function in the solution of the transverse modes in a cylindrical waveguide even not satisfying the imposed boundary conditions. The latter function is necessary to physically and fully describe the formation of the propagating transverse modes focusing attention on the physics and using mathematics as a tool \cite{JEGO}.

In this paper, a detailed physical analysis of the formation of the propagation transverse modes in planar dielectric waveguides using the mathematical-physics approach is presented. Typically, the physical phenomena involved in the description of modes in planar waveguides, are treated in such a way that the balance between the mathematical methods used and the physical constraints of the problem is tipped towards the former, leading to a somewhat unsatisfactory physical description of the formation of the modes in planar dielectric waveguides. Their physical interpretation has been mostly focused on the even modes, which are the real part of the solution, leaving out its imaginary part \cite{Marcuse,Ghatak}. To the best of our knowledge, there has never been an attempt of using the mathematical-physics approach for the physical interpretation of the odd modes.

By doing so, we demonstrate physically that, at wavelength scale, the pure stationary mode in the planar dielectric waveguide is described by the cosine function while the sine function yields a quasi-stationary periodic mode. 

\section{Mathematical-Physics Picture}
\label{Subsec:M-PhysicsPicture}
The vectorial analysis of propagating modes in waveguides is, in general,
very complicated so that a simpler approach is often desirable. In this
context, even the fairly simple case of the two-dimensional refractive index
distributions that occur in channel guides requires elaborated numerical
simulations. Nevertheless, if the index difference $\Delta n$ forming the
guide is small, a scalar approximation can be used instead\cite{Lee}. Bearing this in mind, from now on we will consider only TE polarization for ease of treatment. This plane wave is propagating in a medium with a refractive index $n_{2}$ and impinges on a medium with index $n_1$ such that $n_{1}<n_{2}$. Its direction of propagation is defined by the vector $\overrightarrow{k_{i}}$, as shown in Fig.\ref{fig:Slab}. At the interface this incident wave generates a reflected wave that propagates in the direction of the vector $\overrightarrow{k_{r}}$. At the same time, in the medium of index $n_1$, a transmitted wave is generated, traveling in the direction of the propagation vector
$\overrightarrow{k_{t}}$.

\begingroup
 \centering
  \includegraphics[width=77mm]{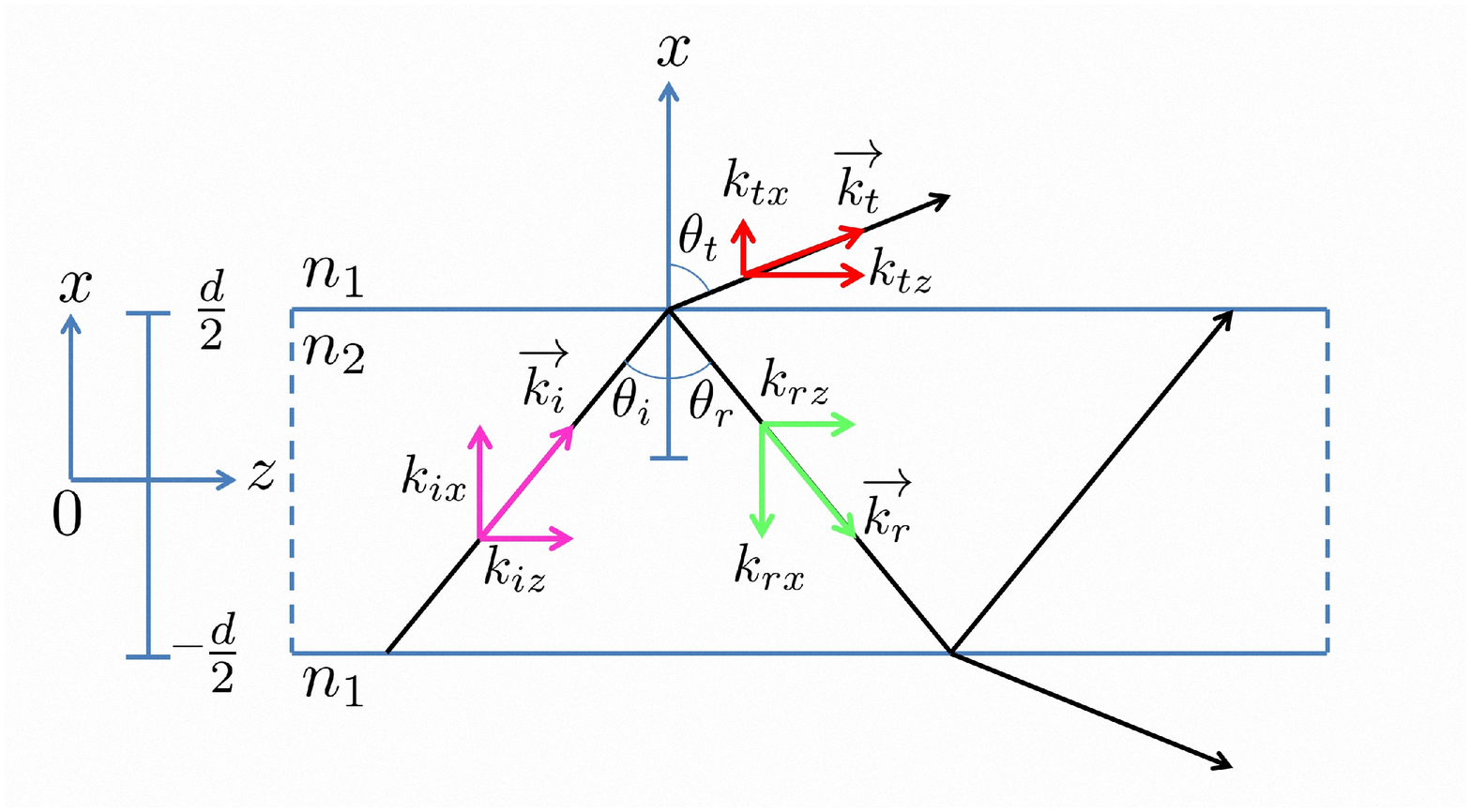}
  \captionof{figure}{Planar waveguide: incident, reflected, and transmitted waves.}
  \label{fig:Slab}
\endgroup

The propagation characteristics can be determined from the geometry shown in Fig. \ref{fig:Slab}. It is straightforward
to note that the following relationships hold for the incident, reflected and transmitted waves:
\begin{equation}%
  \begin{array}
    [c]{cc}%
    k_{ix}=k_{0}n_{2}\cos\theta_{i}, &
    k_{iz}=k_{0}n_{2}\sin\theta_{i},\\ 
    k_{rx}=-k_{0}n_{2}\cos\theta_{r}, &
    k_{rz}=k_{0}n_{2}\sin\theta_{r},\\ 
    k_{tx}=k_{0}n_{1}\cos\theta_{t}, &
    k_{tz}=k_{0}n_{1}\sin\theta_{t}, %
  \end{array}
  \label{ComponentesConIndice}%
\end{equation}
where $k_{0}=2\pi/\lambda$ is the wavenumber in vacuum, with $\lambda$ being the wavelength of the
electromagnetic wave considered. From Eq.
(\ref{ComponentesConIndice}), in particular from the expressions on the right and imposing conservation of tangential momentum, we obtain Snell's law \cite{Mooney}. Those components are fundamental for getting the explicit expressions of the traveling wave modes in the waveguide as we will see next.

The electromagnetic wave propagation inside the slab with refractive
index $n_{2}$ is described by the Helmholtz's equation%
\begin{equation}
  \nabla^{2}E + k_{2}^{2}E = 0.
  \label{helmeq}%
\end{equation}
with wavenumber $k_2=k_0 n_2$. The corresponding plane wave solutions
associated to the wavevectors shown in Fig.\ref{fig:Slab} can be
written as
\begin{equation}
  E_{i}(x,z) = A_{2}e^{ i\left(  k_{ix}x-k_{iz}z\right) },
  \label{eq:Eixz}
\end{equation}
where $A_2$ is a constant.  The expression given by Eq.
(\ref{eq:Eixz}) represents the incident plane wave, with wavevector
$\overrightarrow{k_{i}}$ and with one of its orthogonal components
traveling in the upper direction (as depicted in
Fig. \ref{fig:Slab}). Similarly the expression %
\begin{equation}
  E_{r}(x,z) = A_{2} e^{ i\left(  k_{rx}x-k_{rz}z\right)  },
\end{equation}
corresponds to the reflected plane wave with wavevector
$\overrightarrow{k_{r}}$, whose transverse component travels in the
downward direction, as shown in Fig. \ref{fig:Slab}. Since we know that
$\theta_r=\theta_i$, then we conclude that $k_{rx}=-k_{ix}$ and
$k_{rz}=k_{iz}=k_{z}$.

Let us recall that we are assuming that $n_{2}>n_{1}$ and thus from the direct application of Snell's law we have that there exists an angle $\theta_c$ at which there will be total internal reflection, i.e., the incident field will be totally reflected. This angle is determined by
\begin{equation}
  \theta_{i} = \theta_{c}=\arcsin\left(
    \frac{n_{2}}{n_{1}}\right).
  \label{AngCrit}%
\end{equation}
At this incident angle, the transmitted angle becomes
$\theta_{t}=\frac{\pi}{2}$ and the transmitted wave transforms into an evanescent wave in the transverse direction but with a propagating component along the interface. This is known as an electromagnetic surface or inhomogeneous wave\cite{Corson}. The incident angle at which this happens is called the critical angle, and it is given by Eq. (\ref{AngCrit}). Whenever the incident wave impinges on the interface at angles $\theta_{i}>\theta_{c}$, total internal reflection will take place.

\begingroup
 \centering
  \includegraphics[width=7.7cm]{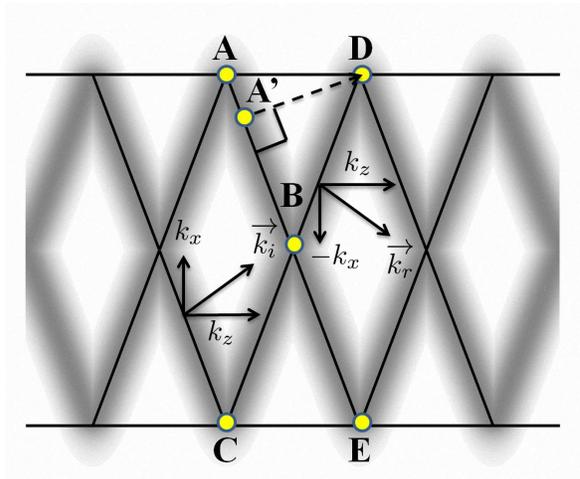}
  \captionof{figure}{Representation of plane counter-propagating waves which are solutions to the Helmholtz equation in a slab. They generate a cosine type solution.}
  \label{fig:Slab_Corson}
\endgroup

Let us now analyze the behavior of the transmitted wave:
\begin{equation}
  E_{t}(x,z) = A_{1} e^{ i\left( k_{tx}x-k_{tz}z\right) },
\end{equation}
where $A_1$ is a constant.
From the geometry of the transmitted wave vector we get, by Snell's
law, that $k_{tx}=\sqrt{k_{1}^{2}-k_{z}^{2}}$, with
$k_{1}=k_0 n_1$, where we have used the simplified notation:
$k_{tz}=k_z$. Concentrating on the case when the incident angle is
greater than the critical angle, $\theta_{i}>\theta_{c}$, after using
Eq. (\ref{ComponentesConIndice}) the expression for $k_{tx}$ can
be written as%
\begin{equation}
  k_{tx} = ik_{0}\sqrt{n_{2}^{2}\sin^{2}\theta_{i}-n_{1}^{2}}=
  i\kappa_{tx}. 
  \label{Dispersion_Eq}
\end{equation}
We would like to point out that the complex exponential becomes a decaying exponential in the transverse direction
and the transmitted wave is described by
\begin{equation}
  E(x,z) = A_{1} e^{-\kappa_{tx}x}e^{-ik_{z}z}.
  \label{outguide}%
\end{equation}
At this point we want to highlight the result given by
Eq. (\ref{Dispersion_Eq}): at the interface the wave number becomes imaginary so that outside the waveguide the ``transmitted'' field is an evanescent wave \cite{Corson}; refer to points A, D, C and E in Fig. \ref{fig:Slab_Corson}. In what follows, we simplify the notation such that $k_{ix}$ is denoted by just $k_{x}$ and $\kappa_{tx}$ by $\kappa_{x}$.

As the transverse electric wave field is reflected from the slab's surface, we will obtain transverse electric waves \cite{Cleveland}. Due to this fact, there is a phase shift of $\pi$ in the reflected waves with respect to the incident wave. This gives rise to up-going and down-going plane waves as shown in Fig. \ref{fig:Slab_Corson}, in which the lines within the waveguide represent the wavefronts of the guided plane waves. The modes of the waveguide will be formed once the appropriate conditions are met in order to have a sinusoidal stationary wave \cite{Corson}.

Such sinusoidal modes are represented by
\begin{equation}
  E(x,z)=\left\lbrace
    \begin{array}{l}
      \cos\left(  k_{x}x\right)  e^{-ik_{z}z} \\
      \sin\left(  k_{x}x\right)  e^{-ik_{z}z}. \\
    \end{array}
  \right.
  \label{SolMF}
\end{equation}
We would like to highlight an important mathematical remark usually
ignored within this context: from Euler's relation
$\exp(\pm i\theta)=\cos\theta\pm i\sin\theta$, we notice that the
cosine function is easily obtained from the sum of a complex
exponential and its conjugate. However, for the case of the sine
function, we must subtract the complex exponential with negative
argument from the one with positive argument. In terms of our physical
system, within the slab there is a superposition of fields; in other words, the
sum of plane waves. This implies that we have to rewrite
Eq. (\ref{SolMF}) to reflect this fact, and thus we obtain that
\begin{equation}
  E(x,z)=\left\lbrace
    \begin{array}{l}
      \frac{1}{2} (  e^{ik_{x}x}+e^{-ik_{x}x})  e^{-ik_{z}z}\\
      \frac{1}{2}e^{-i\frac{\pi}{2}}\left( e^{ik_{x}x}+e^{-i\left[ \pi
            +k_{x}x\right]  }\right)  e^{-ik_{z}z}.\\
    \end{array}
  \right.
  \label{SolFM}
\end{equation}
The second expression implies that in order to have a transverse stationary mode described by the sine function, the down-going wave must be delayed (or advanced) with respect to the up-going wave by a phase shift of $\pi$. This is a remarkable result that is not usually discussed in the general literature on the subject and one that has a profound effect in the understanding of the modes of a planar waveguide as will shall see in Section \ref{subsec:PM_planar}.

The solutions in Eq. (\ref{SolFM}) tell us that, physically, the cosine and sine transverse stationary wave modes inside the waveguide are the result of the sum of counter-propagating components of the transverse traveling plane waves. It is important to remark that the incoming waves are continuously entering and propagating inside the waveguide instead of just one single wave, as shown in Fig. \ref{fig:Slab_Corson}.

\section{Modes created by Propagating waves}
\label{subsec:propag}
Let us now address the problem from the mathematical-physics (m-physics) point of view at the wavelength scale. Suppose that we have a slab of width $d$ and infinite longitude, as shown in Fig. \ref{fig:Slab}. Its physical-mathematics solutions are then given in the core by Eq. (\ref{SolMF}) and in the cladding by Eq. (\ref{outguide}).

By solving the Helmholtz equation (Eq. (\ref{helmeq})) with the separation of variables method in Cartesian coordinates we obtain Eqs. (\ref{SolMF}), which correspond to a purely mathematical
solution. However, if we solve the problem bearing in mind its underlying physics (m-physics), we end up with the solution given by Eqs. (\ref{SolFM}), which represent counter-propagating traveling waves reflecting at both interfaces. It should be clear by now that understanding the true behavior of the
propagation of plane waves inside a planar waveguide can give us a
better physical interpretation of the problem.

Let us then consider a
plane wave that makes an angle with respect to the $x$-axis, and
propagates along the $z$-axis. Now, the question is, how does this
plane wave gets reflected by an interface? The most common approach to answer this question is through a ray analysis. Following the reasoning stressed out in this paper, we rather study the propagation of the wavefronts of the counter-propagating plane waves that constitute in superposition the guided modes.

\begingroup
 \centering
  \includegraphics[width=7.7cm]{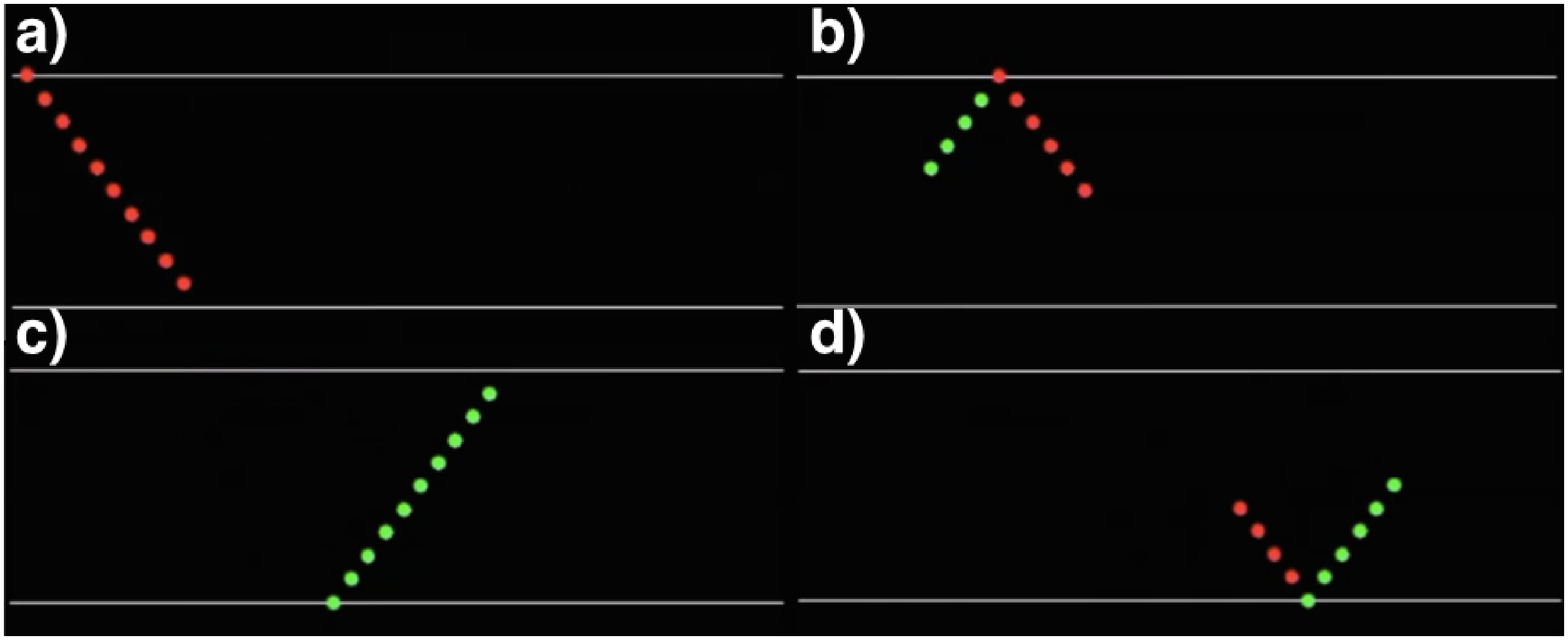}
  \captionof{figure}{The propagation of a wavefront within the slab waveguide. As it propagates part of the wavefront of the incident wave gets reflected several times. The red dots represent sections of the wavefront of the incident wave while the green ones represent the wave front of the reflected wave} 
  \label{fig:Reflection_A_Plane_Wave}
\endgroup

In order to answer this question, and to provide further clarification to the points made above, we have carried out a simulation of the propagation of the wavefronts of the counter-propagating plane waves within the planar waveguide, as shown in Fig. \ref{fig:Reflection_A_Plane_Wave}. For clarity's sake we will
divide the wavefronts into a representative number of sections denoted by dots, where the red and green ones represent a section of the wavefront of the incident and reflected waves respectively. In Fig. \ref{fig:Reflection_A_Plane_Wave} it is shown the propagation of a part of the wavefront of the incident wave. In Fig. \ref{fig:Reflection_A_Plane_Wave}a) it can be observed the incident wavefront whose wave vector lies in the first quadrant of the cartesian plane, that is, propagates to the right and encounters the upper interface. As it propagates part of it becomes the wavefront of the reflected wave, so that the initial red points become green as shown in Fig. \ref{fig:Reflection_A_Plane_Wave}b). Further propagation results in the configuration shown in Fig. \ref{fig:Reflection_A_Plane_Wave}c), where the wavefront is totally reflected just to encounter the interface and be reflected once more, as illustrated in Fig. \ref{fig:Reflection_A_Plane_Wave}d).

In the above discussion we analyzed the propagation of a section of the wavefront of the incident wave. Now we consider a wave packet that enters the slab waveguide and let us assume that it forms an even guided mode so it is described by the upper solution in Eq. (\ref{SolFM}): at any given time, the incident and reflected waves are in phase and thus create a stationary wave. As expected, the wavefronts of both counter-propagating plane waves create a pattern that does not changes along the waveguide, as shown in Fig. \ref{fig:Cosine_Modes} inside the blue rectangle. As the plane waves propagates within the waveguide, the pattern of wavefronts translates without changes. This might seem that the wave fronts propagate from point A to point D of Fig. \ref{fig:Slab_Corson}, instead of moving from point A' to point D as it actually happens. This confusion could arise due to the fact that the wave front pattern does not show any apparent changes as shown in Fig. \ref{fig:Cosine_Modes}.

\begingroup
 \centering
  \includegraphics[width=7.7 cm]{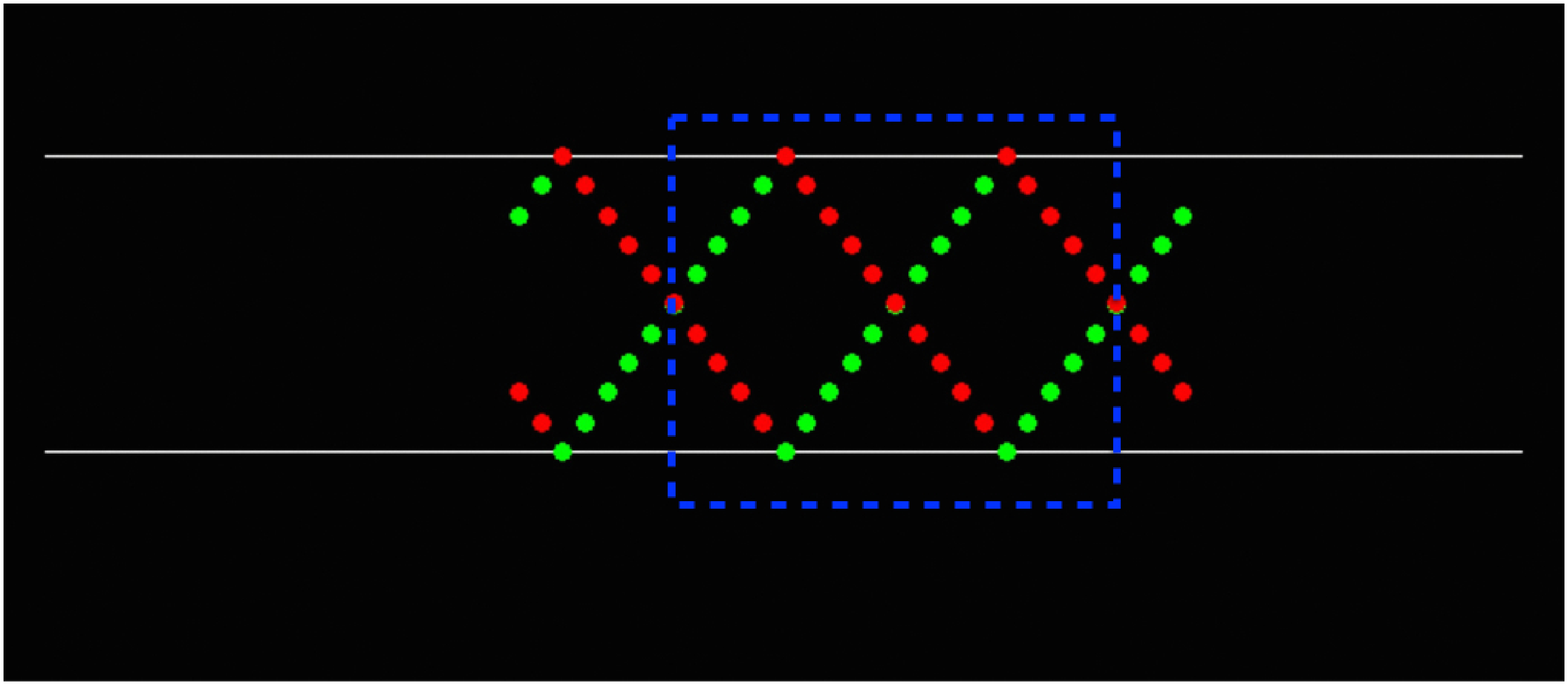}
  \captionof{figure}{Propagation of three plane waves in phase with the reflected
    waves. This pattern represents the standing wave solution of even modes} 
  \label{fig:Cosine_Modes}
\endgroup

From the dispersion relations $\omega/k_{z}=v_{z}$ and
$\omega/k=c$, it can be noted that $k_{z}<k$, and therefore $v_{z}>c$, that is, the phase velocity (velocity of the pattern) is superluminal. By confusing the propagation of the light with the propagation of the pattern, we might conclude that there is a superluminal propagation in a metallic waveguide filled with air. However, this is not the case since the group velocity of the waves inside the waveguide is subluminal.

\begingroup
 \centering
  \includegraphics[width=7.7 cm]{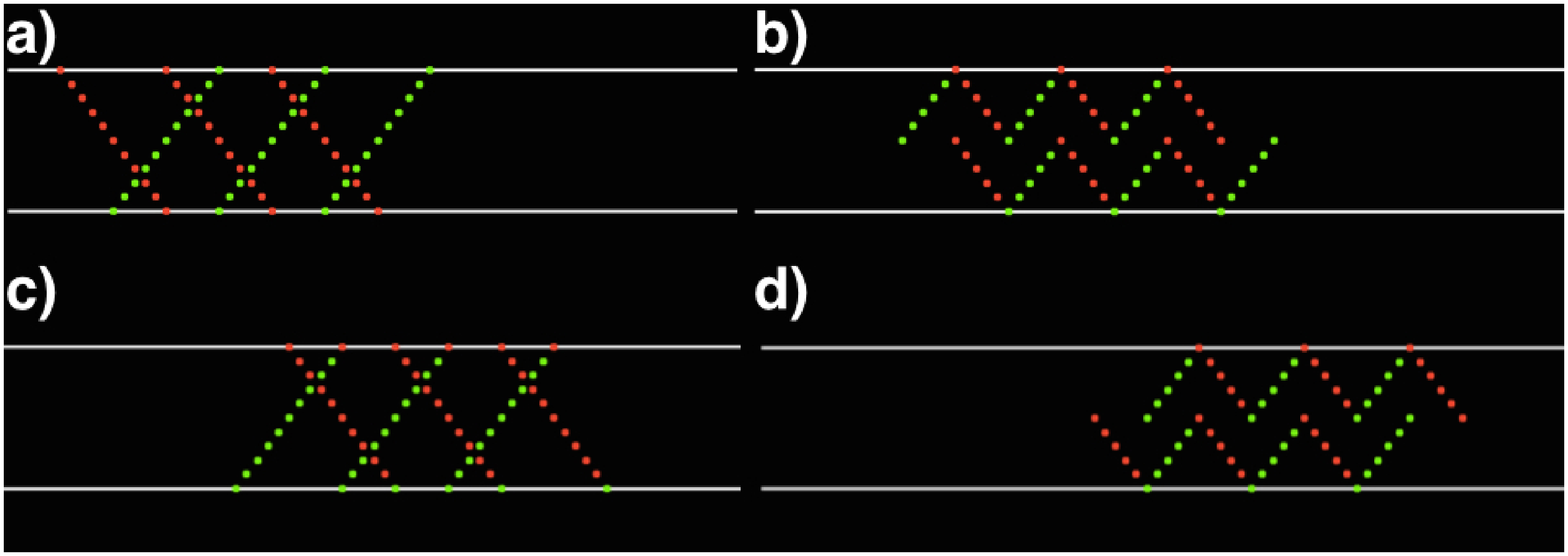}
  \captionof{figure}{The periodic evolution of the wave front pattern of odd modes.} 
  \label{fig:Sine_Modes}
\endgroup

Now assume that the wave packet that enters the waveguide  forms an odd mode so that it is described by the lower solution in Eq. (\ref{SolFM}). Two cases arises: (\textit{i}) the incident plane wave is out of phase with respect to the reflected plane wave by a phase shift of $-\pi$, while in the other case (\textit{ii}) by $+\pi$. In any of those cases a periodic wave front pattern during propagation is created along of the z-axis as shown in Fig. \ref{fig:Sine_Modes}. We think that it cannot be considered as stationary since it does not translates along the $z$-axis without change as shown in Fig. \ref{fig:Sine_Modes}. Therefore we claim that the odd modes are periodic quasi-stationary modes because their wave front pattern recovers its initial shape only after a phase shift of $2\pi$ along the $z$-axis as shown in Fig. \ref{fig:Sine_Modes}. For case (\textit{i}), the wave front pattern follows the order a)-b)-c)-d)-a), while for case (\textit{ii}) the order is c)-d)-a)-b)-c).

The plane waves conforming guided modes in a planar slab waveguide generate stationary modes in terms of cosines and periodic quasi-stationary modes in terms of sines. It is worthy to note that the m-physics approach does not alter the mathematical result, but rather gives a better description of the physical phenomena within the planar waveguide.

In our analysis we have not considered the
  Goos-H\"anchen shift. However, if we include it, it does not change
  the main physical aspect of our approach \cite{Kogelnik}.
  
\section{Physical-Mathematics Picture}
\label{subsec:PM_planar}

In textbook literature it is common to introduce the solutions inside and outside planar waveguides by a purely mathematical approach. These solutions depend on the width and the refractive indices of the waveguide, and are given by
\begin{equation}
  E(x,z)=\left\lbrace
    \begin{array}{l}
      A_{1}e^{-\kappa_{tx}x} \\
       A_{2}\left[ \begin{array}{l}
      \cos\left(  k_{x}x\right) \\
      \sin\left(  k_{x}x\right)
      \end{array}
      \right] \\
      \pm A_{1}e^{\kappa_{tx}x}
    \end{array}
    \right\rbrace e^{-ik_{z}z}
    \begin{array}{ccr}
      x & > & d/2\\
      \mid x \mid & \leq & d/2 \\
      \\
      x & < & -d/2.
    \end{array} 
  \label{SOLNEW}
\end{equation}
It is very important to say that these solutions can be obtained by simply solving the Helmholtz's equation with separation of variables and imposing continuity conditions, that is, by means of purely mathematical instead of physical considerations.

In some textbooks attempts have been made to explain the physics of plate-parallel (or planar-mirror) waveguides. However in some cases this has not been done in the best way. For example in Hayt's textbook we find the following discussion for the explanation of the sine modes \cite{Hayt}: ``The minus sign in front of the second term arises from the $\pi$ phase shift on reflection", in regards to the lower solution in Eq. (\ref{SolFM}) for a parallel-plate waveguide. A reasonable question that can be posed is: why is the minus sign not present in the cosine solution, given that there is also a reflected wave? In other words, even modes cannot be explained in these terms since the $\pi$ phase shift is not present in this case as can be observed in the upper solution in Eq. (\ref{SolMF}), and therefore one of the counter-propagating plane waves conforming the even modes cannot be considered to be the reflected wave of the other one.

This argument holds for both parallel-plate and dielectric slab waveguides. We chose this example since, to the best of our knowledge, there has not been a physical explanation of odd guided modes in dielectric slab waveguides. A simple but important question like this enables us to put into question the physical arguments presented in those textbooks.

Another problem we have noticed bears with the fact that the analysis of guided modes is mainly done for the even ones rather than for both of them, i.e., the even and odd modes. The easy, but dangerous explanation, is that the physical analysis of odd modes is skipped because the authors probably assume wrongly that the odd solutions have a similar behavior to that of the even ones.

In contrast to the remarks above, in Sections \ref{Subsec:M-PhysicsPicture} and \ref{subsec:propag} we have given a complete physical explanation of these solutions.

\section{Conclusions}
We demonstrate that, when using the mathematical-physics approach is possible to easily visualize the transverse solutions inside a slab waveguide as the sum of counter-propagating traveling plane waves, i.e., their modes are composed transversely by standing waves. We have demonstrated that these waves, physically, can only generate stationary modes in terms of cosines and periodic quasi-stationary modes in terms of sines. For the cosine profiles, the incident and reflected plane waves are in phase, and it is clearly possible to see that the standing wave is formed. Inside the planar waveguide, the sine profiles have a  phase shift between the reflected and the incident waves. This phase shift generates a quasi-stationary periodic mode propagating inside the waveguide.

\section*{Funding}
Consejo Nacional de Ciencia y Tecnolog\'ia (CONACYT) (235164).

\end{multicols}
\begin{multicols}{2}






\end{multicols}
\end{document}